\documentclass[aps,showpacs,preprintnumbers,twocolumn,prl]{revtex4} 
\usepackage{graphicx,color,epsfig,rotate}
\usepackage{amsbsy,amssymb,amsmath,bm}
\usepackage{dcolumn}
\usepackage{bm}
\usepackage{umoline}
\begin{document}


\title{Avalanches and disorder-induced criticality in artificial spin ices}

\author{Gia-Wei Chern, C. Reichhardt, and C.~J. Olson Reichhardt}
\affiliation{Center for Nonlinear Studies and Theoretical Division, Los Alamos National Laboratory, Los Alamos, NM 87545, USA}  

\date{\today}

\begin{abstract}
We show that both square and kagome artificial spin ice systems exhibit 
disorder-induced nonequilibrium phase transitions,  with power law avalanche distributions 
at the critical disorder level. The different nature of geometrical frustration in the two lattices 
produces distinct types of critical avalanche behavior. For the square ice, the avalanches 
involve the propagation of locally stable domain walls separating the two polarized ground states, and the scaling collapse agrees with 
an interface depinning mechanism.  In contrast, avalanches in the fully frustrated kagome ice
exhibit pronounced branching behaviors that resemble those found in directed percolation. The kagome ice 
also shows an interesting crossover in the power-law scaling of the avalanches at low disorder. 
Our results show that artificial spin ices are ideal systems in which to study nonequilibrium critical point phenomena.
\end{abstract}


\maketitle
When collective interacting systems are driven in  the presence of disorder,
the dynamics often takes the form of intermittent bursts of avalanches of activity. In many cases, such as in earthquakes \cite{carlson94}, 
intermittent flow of dislocations \cite{miguel01}, flux avalanches in superconductors \cite{nowak97}, and crackling or Barkhausen noise in 
magnetic systems \cite{sethna01,sethna93}, the avalanche events exhibit scale-free power law distributions, 
indicative of critical phenomena \cite{sethna01}.
The avalanche behavior in some of these systems can be understood using the concept of self-organized critically, 
in which the driven system tunes itself to a critical state so that, unlike 
in equilibrium systems, there is no tuning parameter \cite{bak87}.  An alternative approach is to assume that
a true critical point is present, and that the amount of disorder must be tuned 
at that point to generate avalanche distributions  that obey a power law at all length scales \cite{sethna01,perkovic95}. 

The nonequilibrium random field Ising model (RFIM) \cite{sethna01,sethna93,perkovic95,spasojevic11} is 
the most important system to which the critical point approach has been applied. In this model, as the external field changes, the reconfiguration of the
spin arrangements occurs in the form of avalanche events. At low disorder strength below the critical point,
the dynamics is dominated by large system-spanning events, while at the critical amount of disorder $r_{c}$, 
there are avalanches on all size scales with no cutoff in the distribution.  As the disorder strength is increased above $r_c$, a cutoff appears that
drops to smaller avalanche sizes for larger disorder. In this picture, the critical fluctuations are not associated with
self-tuning but with an actual critical point that has a tuning parameter; however, in many cases the avalanche behavior persists a considerable
distance away from $r_c$, making a scaling collapse of the avalanche distribution one of the best methods for revealing 
the existence of a critical point \cite{perkovic95}.  Since the introduction of the RFIM, the model and variations of it
have been applied to a number of other systems beyond magnetic spins. 
Although numerical models can access microscopic spin configuration  information, this is generally not possible
in experimental systems, particularly for spin systems.

A recent example of systems in which microscopic effective spin information is directly accessible experimentally is artificial spin ices \cite{wang06,nisoli13}. 
Here, arrays of interacting nanomagnets are arranged in a geometry that  models two-dimensional (2D) spin ice, with the magnetic moment of the individual 
nanoisland acting as an effective macroscopic spin.  Artificial ice studies have generally focused on proving that the system exhibits what are 
termed ``spin ice rule'' obeying states  as well as the frustration effects that are found in actual spin ices \cite{moessner06}. 
The excitement generated by these systems arises from the fact that the effective spin degrees of freedom are on a 
size scale that can be directly imaged, giving access to the defects, domains, orderings, and dynamics of the
artificial ice. Experiments have considered square ice in which four nanoislands meet at each vertex \cite{wang06,Marrows,kapaklis12}, 
illustrated in Fig.~1(a), as well as kagome ice (or honeycomb array) \cite{qi08,ladak10,mengotti11,zhang13}, shown in Fig.~1(b). 
Beyond magnetic systems, artificial spin ice arrangements have been studied in soft matter systems \cite{libal06,han08}  
and in superconducting arrays \cite{libal09,latimer13}.

Recent experiments also captured the dynamics, e.g. collective motions of defects and avalanche behavior in artificial ice arrays.
Ladack {\it et al.} imaged the motion of topological defects in connected kagome arrays and constructed a magnetic  hysteresis loop \cite{ladak10}. 
Mengotti {\it et al.} studied artificial 2D kagome  ice systems at room temperature
and observed sudden collective events of effective spin reconfiguration \cite{mengotti11}.
The distribution of these avalanche events $D(s)$ was exponential rather than power law, in agreement with
Monte Carlo studies for this system performed at three disorder values. It was argued in Ref.~\cite{mengotti11} that
since the avalanches formed one-dimensional (1D) structures that differed from avalanches in the 2D RFIM, a power law behavior of the
distribution would not be expected; however,  the absence of power-law behavior could be due to 
boundary effects in the finite clusters or the fact that the amount of disorder  is not close to the critical point.

\begin{figure}
\includegraphics[width=.9\columnwidth]{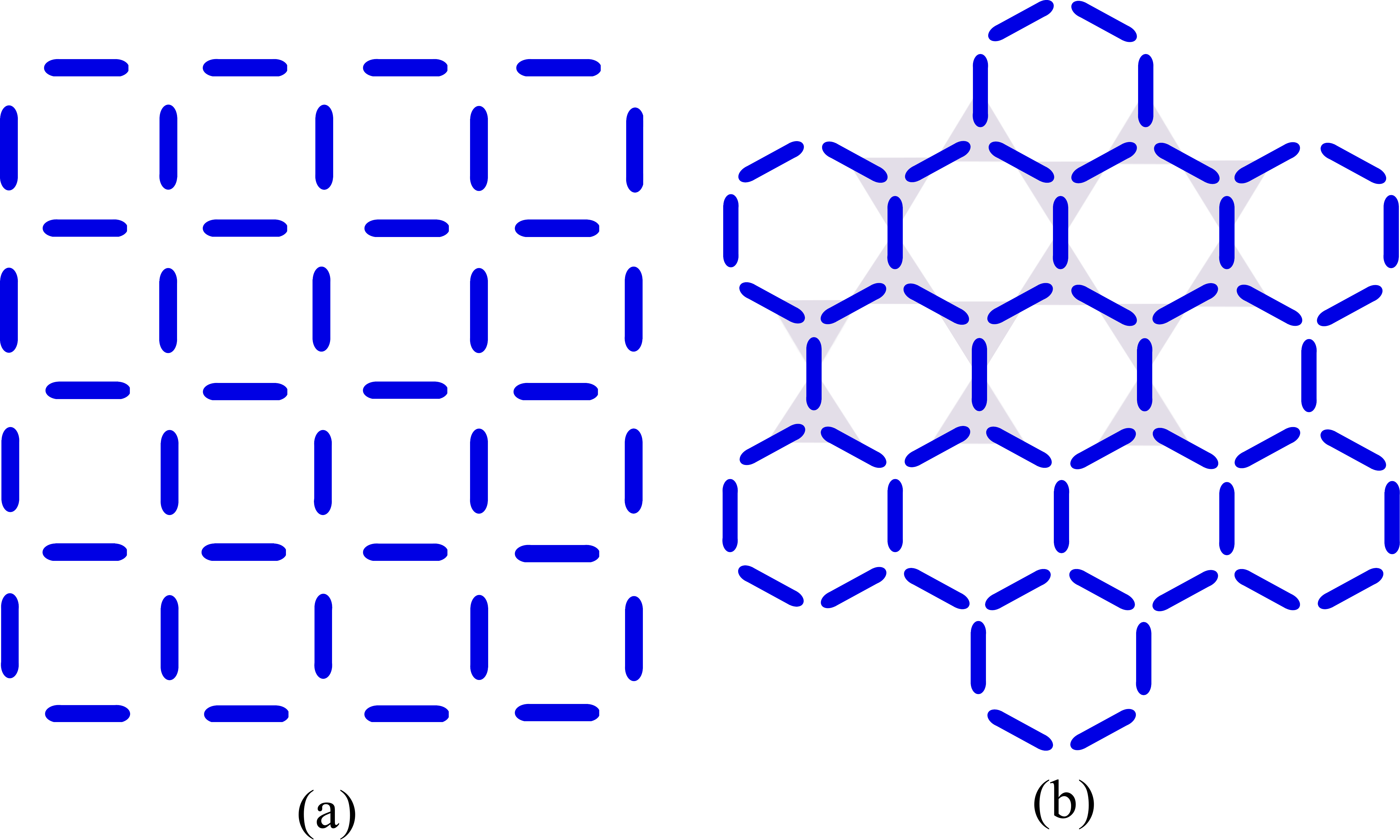}
\caption{(Color online)  2D arrays of nanomagnets for (a) square ice and (b) kagome ice.
The finite clusters used in our simulations possess $C_4$ and $C_6$ 
symmetry in the square and kagome cases, respectively.
The boundary vertices in the square ice cluster have coordination number $z = 3$ at the edge and $z=2$ at the corners,
while the boundary vertices in the kagome ice cluster always have coordination $z=2$ (armchair geometry).
\label{fig:array} 
}
\end{figure}

In this Letter we study the avalanche dynamics in spin ice arrays 
consisting of {\it disconnected} islands, as realized in several recent 
experiments~\cite{wang06,mengotti11,zhang13}. We find that magnetization reversal in these systems proceeds via 
spin reconfiguration avalanches. By studying the dependence of the magnetization dynamics on disorder, which is always present in real systems, 
we show that the spin avalanches exhibit different behaviors depending on the amount of the disorder. In particular, there exists a critical disorder 
$r_c$ at which avalanches of all length scales up to the finite lattice size are observed. We also find that 
the critical behaviors of the avalanches in the two geometries are related to the type of topological defects present in the spin ice systems.

We consider finite square and honeycomb arrays that possess $C_4$ and $C_6$ symmetry, respectively, as shown in Fig.~\ref{fig:array}.  
Quenched disorder is included by varying the coercive switching fields $H_{c,\,i}$ of individual links. 
Specifically, we obtain $\{H_{c,\,i}\}$ from a Gaussian distribution with average $\bar{H}_c$ and standard deviation $\sigma_{H_c}$, and we 
characterize the strength of the disorder by the ratio $r \equiv \sigma_{H_c} / {\bar{H}_c}$.
The magnetization of an individual single-domain nanoisland is described by an effective Ising spin, 
$\mathbf m_i = \sigma_i m_s \hat{\mathbf e}_i = \pm m_s \hat{\mathbf e}_i$, where $m_s$ is the saturation 
magnetization and $\hat{\mathbf e}_i$ is a unit vector pointing along the spin direction of island $i$. 
Due to the large size of the arrays studied here (number of links $N\sim 10^6$), it is not feasible to perform full scale 
micromagnetic simulations. Instead, we approximate individual nanomagnets 
as point dipoles interacting with each other via dipolar interactions.  We employ a zero-temperature relaxation dynamics for the Ising spins. 
An Ising spin $\sigma_i$ is flipped if the total field acting on it, composed of the external field $\mathbf H$ and the dipolar 
field $\mathbf h^{\rm dip}_i$ from all other islands, exceeds its coercive switching field: 
$\bigl(\mathbf H + \mathbf h^{\rm dip}_{i}\bigr)\cdot \hat{\mathbf e}_i < -H_{c,\,i}$~\cite{moller06,ladak10,budrikis12,pollard12}.
It is worth noting that the dynamics used here is different from that 
for arrays of connected islands~\cite{daunheimer11}, where the 
dipolar interactions are subsumed into Coulomb interactions between 
magnetic charges and the flipping of each nanomagnet is
mediated by the emission and subsequent absorption of domain walls 
by the vertices~\cite{mellado10}.

In our simulations, the nanomagnetic arrays are initially polarized by an external field applied along the high-symmetry diagonal and vertical 
directions for the square and kagome ices, respectively. The external field is then applied in the opposite direction with gradually 
increasing magnitude.  To achieve this, we increase the field in small steps $\Delta H = 0.01 {\bar H}_c$. At each field increment, we first search 
for the flippable spin satisfying the condition $\mathbf H^{\rm tot}_i \cdot\hat{\mathbf m}_i < -H_{c,\,i}$. 
Generally there is at most only one flippable spin in the so-called adiabatic limit.  We invert this spin and compute the corresponding change 
in the dipolar fields for all other spins within a radius of 50 lattice constants to approximate the long-range dipolar interactions. 
A system-wide search for flippable spins is conducted and the whole process is repeated until no flippable spins can be found.

\begin{figure}
\includegraphics[width=0.99\columnwidth]{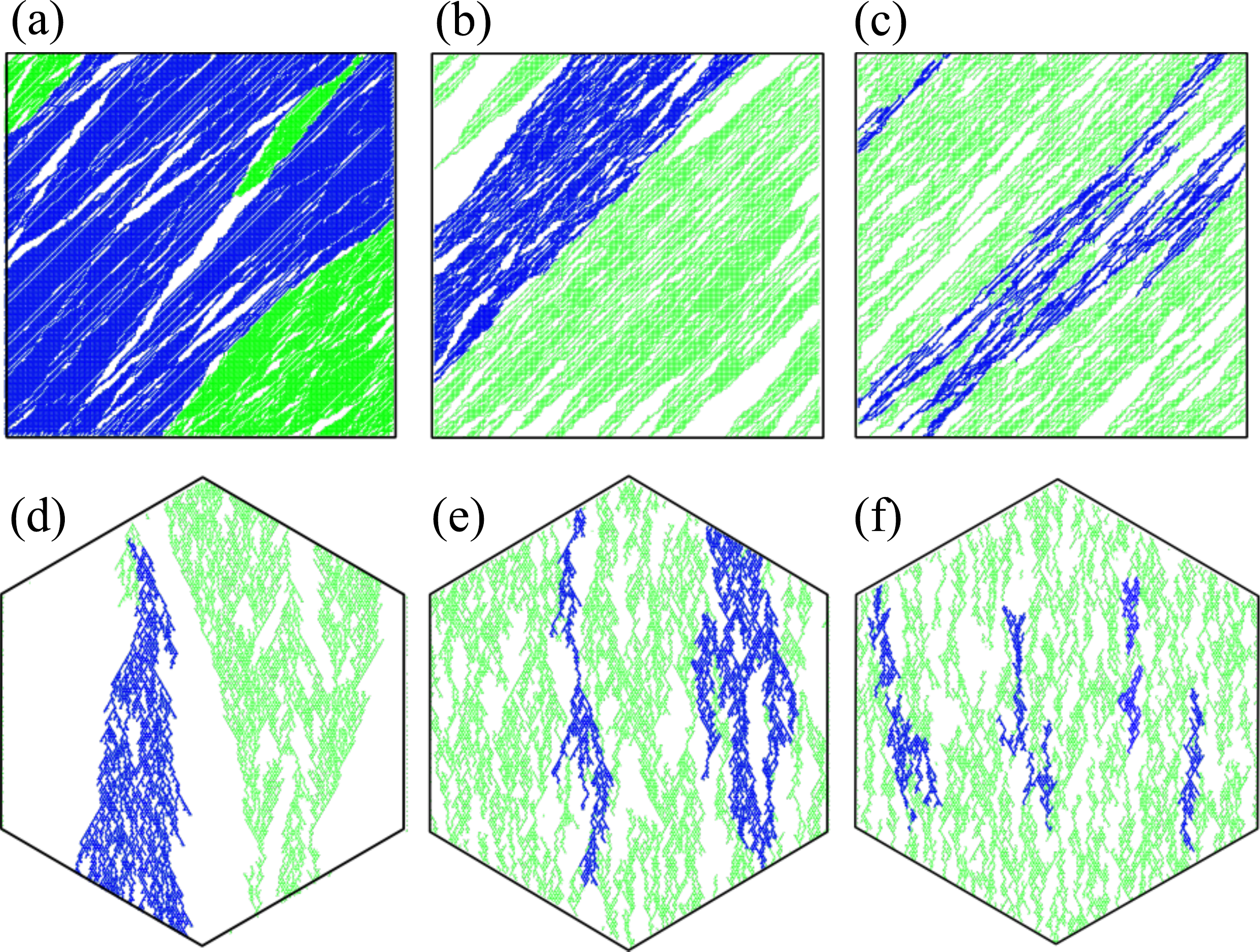}
\caption{(Color online) Snapshots of the spin configurations for square and 
kagome spin ice arrays. The green area denotes the inverted spins while the 
blue area indicates some of the avalanche clusters. The disorder parameter 
is $r =$ (a) 0.012, (b) 0.018, and (c) 0.023 for the square ice snapshots 
and $r =$ (d) 0.06, (e) 0.10, and (f) 0.12 for the kagome ice.
\label{fig:kagome1} 
}
\end{figure}

{\em Square spin ice}. We first consider square ice arrays, as shown in Fig.~1(a), 
with an external field applied along the $45^{\circ}$ symmetry direction.  
Representative avalanche clusters in the regime of weak, critical, and strong disorder are shown in Fig.~2(a)--(c).
For weak disorder, there are a small number of very large events that sweep through the 
length of the system, as shown in Fig.~2(a), while at strong disorder, large avalanches are cut off
at a length scale that decreases with increasing disorder strength; see Fig.~2(c). In the vicinity of a critical
disorder $r_{c}$, avalanches of all sizes occur, including some as large as the system size.

\begin{figure}
\includegraphics[width=0.99\columnwidth]{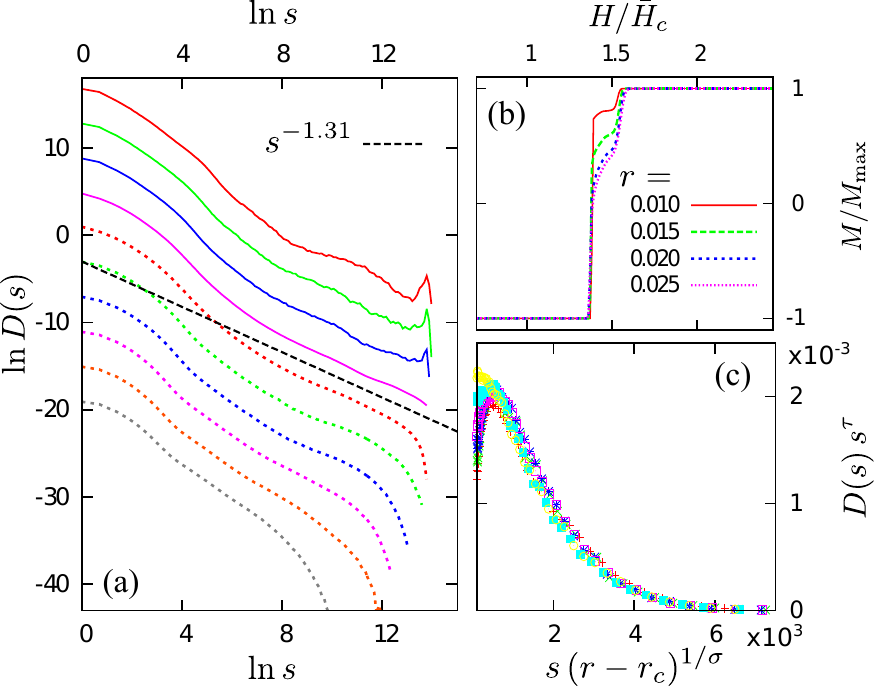}
\caption{(Color online) (a) Avalanche size distributions $D(s)$ during magnetization reversal in a square spin ice consisting of $N = 1281600$ spins. The different curves correspond to disorder level $r = 0.011$, 0.012, 0.013, 0.014, 0.015, 0.016, 0.018, 0.022, 0.026, 
and 0.030 (from top to bottom). The distribution exhibits a power-law $D(s) \sim s^{-1.31}$ in the vicinity of the critical disorder $r_c \approx 0.0145$. (b) Magnetization $M/M_{\rm max}$ vs applied field $H/\bar{H}_c$ at various disorder strengths $r$ in a square ice. Here $M_{\rm max} = \sqrt{2}\, N M$ is the maximum magnetization along the diagonal direction. (c) Scaling collapses of $D(s)$ with parameters $\tau = 1.32$ and $1/\sigma =  0.72$. 
\label{fig:square1} 
}
\end{figure}

The avalanche size distribution $D(s)$ shown in Fig.~\ref{fig:square1}(a) exhibits two distinct behaviors depending on the level of disorder. For weak 
disorder such as $r \sim 0.011$ to $0.013$, a peak in $D(s)$ at the largest $s$ indicates that the dynamics is dominated by large events, consistent with the picture of a supercritical regime described above. The corresponding magnetization curves $M(H)$ shown in Fig.~\ref{fig:square1}(b) are characterized by a pronounced jump in $M$, indicating a system-size avalanche event. On the other hand, for larger disorder such as $r > 0.015$, the large avalanches are cut off at a characteristic size $s_m$ that decreases with increasing $r$. The magnetization in Fig.~\ref{fig:square1}(b) also shows a smooth variation with field in 
this subcritical regime.

Close to a critical value of $r_{c} \approx 0.0145$, the avalanche distribution has a power law form, $D(s) \sim s^{-\tau}$ with an exponent $\tau = 1.31$, 
as indicated by the dashed line in Fig.~3(a). A more accurate method for determining $r_{c}$ and $\tau$ is to perform a data collapse by 
plotting the scaled distribution $D(s)s^{\tau}$ as a function of $(r - r_c)^{1/\sigma}s$, indicating a power-law divergence of the cut-off 
avalanche size $s_m \sim (r - r_c)^{-1/\sigma}$. Fig.~\ref{fig:square1}(c) shows the scaling collapse with $\tau = 1.31$, $1/\sigma  = 0.72$, and $r_{c} = 0.0145$. 
The critical avalanche behaviors observed here can be understood in terms of 
propagating domain walls separating the two polarized states~\cite{durin00}. 
Since the so-called type-II vertices have a net magnetization pointing along the diagonal~\cite{wang06,nisoli13}.
these polarized states, in which all vertices are in one of the two type-II ice-rule obeying configurations, 
are ground states of the square ice in the presence of a magnetic field pointing along the 45$^\circ$ direction, 
The interface between two domains of opposite polarity is composed of type-I vertices (symmetric 2-in-2-out configurations~\cite{wang06}) 
that are locally stable, lowest-energy states of the dipolar interaction.
For square spin ice systems, domain walls are known to form for weak or intermediate disorder, 
but are destroyed in strongly disordered samples \cite{libal09,Marrows,budrikis12,budrikis12b}.

\begin{figure}
\includegraphics[width=0.99\columnwidth]{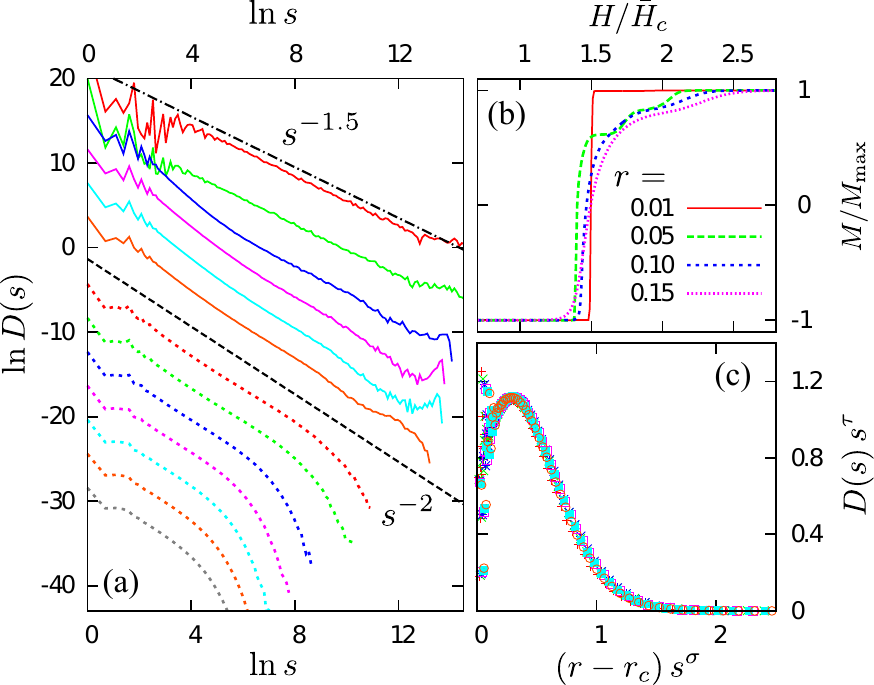}
\caption{(Color online) (a) Avalanche size distribution $D(s)$ during magnetization reversal in a kagome spin ice consisting of $N = 2420112$ spins. The different curves correspond to a disorder level $r = 0.02$, 0.03, 0.05, 0.07, 0.08, 0.09, 0.1, 0.105, 0.1, 0.115, 0.125, 0.135, and 0.145 (from top to bottom). In the vicinity of the critical disorder $r_c \approx 0.107$, we find power law behavior $D(s) \sim s^{-2}$. This exponent gradually changes from $\tau = 2$ to 1.5 in the supercritical regime below $r_c$. (b) Magnetization $M/M_{\rm max}$ vs applied field $H/\bar{H}_c$ at various disorder strengths $r$ in a kagome ice. (c) Scaling collapses of $D(s)$ with parameters $\tau = 2.06$ and $\sigma = 0.63$. 
\label{fig:kagome1} 
}
\end{figure}

\begin{figure}
\includegraphics[width=0.92\columnwidth]{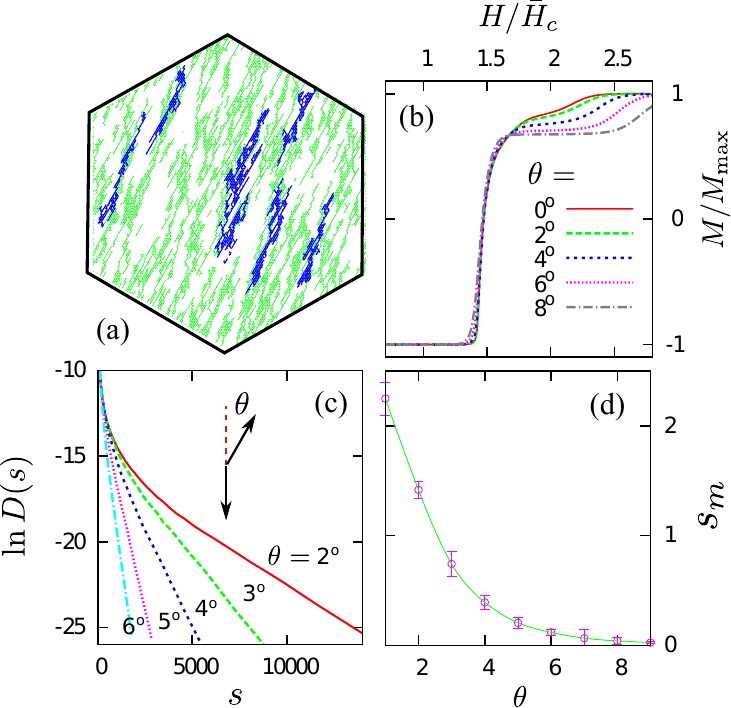}
\caption{(Color online) (a) A snapshot of the spin configurations with an applied field angle $\theta = 5^\circ$. The green area denotes the inverted spins while the blue area represents some of the avalanche clusters. (b) Magnetization $M/M_{\rm max}$ as a function of applied field $H$ at different angles. (c) Avalanche size distribution in log-linear scale showing the exponential decaying form $D(s) \sim \exp(-s/s_m)$ at large $s$. (d) The cut-off avalanche size $s_m$ as a function of the field angle $\theta$.
\label{fig:kagome2} 
}
\end{figure}

{\em Kagome spin ice}. We next consider avalanches in the kagome ice arrays. While the avalanche size distribution 
in Fig.~\ref{fig:kagome1}(a) has critical behavior similar to that found in square ice, the geometry of the avalanche clusters is rather different.
Figures~2(d)--(f) show that the growth of the avalanche clusters in kagome ice exhibits a high degree of branching at different levels of disorder. 
The avalanche size distribution $D(s)$ at weak disorder is characterized by either a plateau or a spike feature at the largest 
values of $s$, indicating an excess number of system-size avalanches. In this supercritical regime, such as at $r = 0.01$, 
the change in magnetization $M$ occurs almost in a single step corresponding to the onset of a system-size avalanche, 
as shown in Fig.~\ref{fig:kagome1}(b). Interestingly, we find two distinct scalings in the kagome ice supercritical regime, 
unlike the case of square ice. At $r = 0.02$, the avalanche size distribution 
in Fig.~\ref{fig:kagome1}(a) has a power-law part $D(s) \sim s^{-1.5}$, as indicated by the upper dot-dashed line. As $r$ increases further, the power-law distribution 
crosses over to $D(s) \sim s^{-2.0}$, as shown by the lower dashed line. This crossover phenomenon in $D(s)$ might reflect the reduced 
dimensionality of avalanche clusters with increasing $r$. At low disorder, the avalanche proceeds with many branchings and exhibits a 2D like structure, 
while the increased amount of disorder reduces the branching, producing more 1D-like avalanches.

For larger disorder, the distribution $D(s)$ develops a characteristic cutoff $s_m$ for large values of $s$ which decreases with increasing disorder in the 
subcritical regime. Our simulations show strong evidence for a critical point near $r_c \approx 0.107$ separating the supercritical and subcritical
regimes. The distribution at $r_c$ is best fit by a power law $D(s) \propto s^{-2.0}$. Fig.~2(e) illustrates the avalanche clustering in the kagome system at $r = 0.10$ near the critical point, where the avalanche structures can be as large as the system size, while in Fig.~2(f), above $r_c$ at $r = 0.2$ the avalanches are smaller, exhibit less branching, and have 1D characteristics.  This supports the idea that the experiments of Ref.~\cite{mengotti11} were performed in the strong disorder regime, where 1D avalanche structures are more prevalent.

In Fig.~\ref{fig:kagome1}(c) we show the scaling collapse $D(s)s^{\tau}$ vs $(r-r_c)\, s^{\sigma}$ for the kagome system, with $\tau = 2.08$, $\sigma  = 0.63$, and $r_{c} = 0.107$.  
The 2D RFIM values are $\tau = 2$ and $\sigma = 0.24$, similar to the
kagome ice value of $\tau$ but different from the value of $\sigma$,
indicating that kagome ice avalanches do 
not fall in the 2D RFIM universality class.
It has been suggested that the exponent $\tau \sim 2$ is a result of super-universality for certain class of time-directed avalanches~\cite{maslov95}. In fact, the high degree of branchings and strong unidirectional growth of the avalanches suggest that the kagome system might belong to the universality class of directed percolation.

Finally, we also study the dependence of avalanches on the direction of the reversal field~\cite{mellado10}. The spin-ice array is first polarized by a field in the $-y$ (or 180$^\circ$) direction. The system is then subject to a reversal field $\mathbf H = H (\sin\theta, \cos\theta)$, {\it i.e.}, along a direction which deviates from $+y$ by an angle $\theta$. Our simulation 
results in Fig.~\ref{fig:kagome2} show a pronounced 1D character of the avalanches, even for small angles $\theta \gtrsim 5^\circ$. The avalanche size distribution has a clear exponential tail $D(s) \sim \exp(-s/s_m)$, where the cutoff $s_m$ quickly decreases with increasing $\theta$.

In summary, we have shown that square and kagome artificial spin ice models
with disconnected islands exhibit disorder-induced nonequilibrium phase transitions.
The critical point of the transition is characterized by a diverging 
length scale and the effective spin reconfiguration avalanche sizes
are power-law distributed. For weak disorder, the magnetization reversal is dominated by large avalanche events that span the entire system, characteristic of a supercritical regime. On the other hand, at strong disorder (the subcritical regime) the avalanche distributions are cut off above a length scale that decreases with increasing disorder. For the square ice, we find a power-law 
avalanche size distribution at the critical disorder level with an exponent $\tau \approx 1.31$. The presence of a stable interface separating the two polarized states in square ice arrays suggests that the critical avalanches belong to the universality class of domain-wall depinning transitions. A rather different avalanche geometry is observed in kagome spin ice arrays. In the vicinity of the critical disorder level, the avalanches exhibit a high degree of branching and unidirectionality. Both features 
are consistent with a universality class of directed percolation or invasion models. The kagome system also shows an interesting crossover from $\tau=1.5$ to $\tau=2$ for the power-law part of the distribution in the supercritical regime. 
Our results show that artificial spin ice can be an ideal system in which to study a variety of nonequilibrium critical phenomena where the microscopic degrees of freedom can be accessed directly in experiments.


{\it Acknowledgments}. We thank K. Dahmen, I. Gilbert, H. Katzgraber, R. Moessner, C. Nisoli, O. Tchernyshyov, and P. Schiffer for
helpful discussions and comments.  G.W.C. thanks S. Daunheimer and J. Cumings for sharing 
unpublished results and for collaboration in a related work.

\end{document}